\documentclass[a4paper,12pt]{article}
\usepackage{lipsum}
\RequirePackage[margin=1in,lines=25,pdftex]{geometry}
\usepackage{setspace}
\usepackage{amssymb, booktabs}
\usepackage{amsmath}
\usepackage{amsthm}
\usepackage{tablefootnote}
\usepackage{dsfont}
\usepackage{url}
\usepackage{natbib}
\usepackage[utf8]{inputenc}
\usepackage[T1]{fontenc}
\usepackage[pdftex]{graphicx}
\usepackage{comment}
\DeclareGraphicsExtensions{.png, .pdf, .jpg} 
\usepackage[pdftex, colorlinks, linkcolor=blue, urlcolor=blue, citecolor=blue, breaklinks=true]{hyperref}

\newtheorem{thm}{Theorem}
\newtheorem{lem}{Lemma}
\newtheorem{fol}{Corollary}
\newtheorem{bem}{Remark}

\begin{document}


\begin{center}
 {\large \textbf{Optimal allocation of trials to sub-regions in crop variety testing with multiple years and correlated genotype effects}}
 \end{center}

\begin{center}

{\small
\textbf{Maryna Prus$^1$, Lenka Filov{\'a}$^2$, Hans-Peter Piepho$^1$, Waqas Ahmed Malik$^1$}

\vspace{0.5cm}

$^1$ Biostatistics Unit, University of Hohenheim, Stuttgart, Germany

$^2$ Department of Applied Mathematics and Statistics, Comenius University, Bratislava, Slovakia
}
\end{center}
\vspace{0.2cm}

{\footnotesize
SUMMARY: Plant breeding and variety trials are usually conducted in multiple environments sampled from a defined target population of environments in order to characterize the performance of breeding lines or varieties. When the population is large and heterogeneous, it may be sub-divided into sub-regions or zones according to administrative and agro-ecological criteria. Analysis then focuses on prediction of performance in the individual sub-regions. Modelling the genotype effect in each sub-region as random, information can be borrowed across sub-regions using best linear unbiased prediction based on a suitable variance-covariance matrix for the genotype-zone effects. Here, we consider the important case where kinship of pedigree information is available for the genotypes under test. This information can be integrated into the variance-covariance matrix for genotype-zone effects. The objective we pursue here is to determine the optimal allocation of a fixed budget of trials to sub-regions. This design problem is solved using a combination of theory and explicit equations on one hand and numerical optimization on the other hand. Our proposed novel approach allows obtaining the optimal allocation when the number of genotypes is in the hundreds, a common setting in large plant breeding programs as well as in variety testing for economically important crops.

}

\newpage
\section{Introduction}



In plant breeding and variety testing, candidate varieties are tested at multiple locations and years to assess their performance in relevant agronomic traits such as yield. Analysis of such multi-environment trials (MET) is routinely done by linear mixed models (LMM) with effects pertaining to the factors variety, location and year. To make predictions into the target population of environments (TPE), it is appropriate to model all effects involving location or year as random (\citealt{piepho2024}).

For trialing systems with wide geographical coverage, it is common to subdivide the TPE into sub-regions or zones (\citealt{atl}). In such systems, the stratification of locations into sub-regions can be integrated into the LMM, introducing a sub-region-specific random effect for variety and assuming that effects of the same variety in different sub-regions are correlated. Such a model allows making sub-region-specific best linear unbiased predictions (BLUP) of variety performance borrowing strength across sub-regions.

An important design problem in MET for sub-divided TPE is the allocation of trials to sub-regions. The optimal allocation depends on the variances and covariances of random effects, in particular of the sub-region-specific variety effects. Solutions to this design problem have been proposed by \cite{pru9}, \cite{pru10}, assuming that random effects of different varieties are independent.

In plant and animal breeding it is common to allow for covariance among genotypes due to pedigree relationships (\citealt{hen3}) or genomic kinship (\citealt{meuw}). This type of model allows sharing information across genotypes, improving accuracy of BLUPs. Optimal design under such models has been considered by \cite{gil1} and \cite{cul} among others. As a proof-of-concept, \cite{prus2025computing} considered the optimal allocation of trials to sub-regions in a TPE based on a model with random sub-region-specific variety effects for a hypothetical example with a small number of varieties. For simplicity, that model was for single-year data.  The computational approach proposed in that work is based on the method developed by \cite{harm} using the MISOCP procedure of the OptimalDesign package in R (\citealt{sag}). In practice, the number of varieties is rather larger, typically ranging from small multiples of ten to several hundreds or even over a thousand in early generations of a breeding program. A solution to the optimal allocation of trials to sub-regions in such large MET is as yet lacking. Given the high memory requirements of the MISOCP procedure, the method proposed by \cite{prus2025computing} becomes unfeasible when dealing with a large number of varieties. The purpose of our paper is to close this gap. Here, we investigate optimal allocation of trials to sub-regions for multi-annual experiments using pedigree relationships among genotypes. Moreover, we consider MET spanning multiple years so that year-specific effects need to be added in the model. {For the extended model, we derive the mean squared error (MSE), and we search for the designs minimizing the MSE.
For computation of optimal designs we employ a modified version of the algorithm proposed in \cite{rad} that allows for large numbers of genotypes and imposition of additional linear constraints, such as cost limitations. 

The paper has the following structure: Section~2 describes the experiment and the datasets. 
In Section~3 the LMM used for the design purposes is specified and the MSEs for the BLUPs of the genotype-zone effects are determined. Section~4 presents analytical results for the optimal allocations of trials and the computational approach. Section~5 illustrates the results obtained in Section~4 by real data examples. The paper is concluded by a short discussion in Section~6. All mathematical derivations are deferred to Web Appendices.

\section{Description of the Experiment and the Data}\label{data}


\textbf{Datasets}

We used variance component estimates from a zoned MET conducted within the All-India Coordinated Maize Improvement Program (AICMIP) (\citealt{kle}). As no marker were available for this MET, we employed kinship matrices for maize from three different sources, comprising 31, 80 and 698 genotypes. In each case, the kinship matrix was computed according to the VanRaden method (\citealt{vanr}), meaning that the genetic variance components in the phenotypic MET dataset can be used at face value in combination with the kinship matrices (\citealt{feld}).

\textbf{Phenotypic dataset}

The AICIMP trials span five agroecological zones for maize in India. The data comprise the years 1995 through 2006, and four maturity groups, which were analyzed separately. The linear mixed model used for analysis has the same fixed and random effects as our model (1) (see the next section). We use the variance components reported in \cite{kle}.

\textbf{Marker dataset 1}

A multi-site experiment involving 36 field experiments (each defined as a combination of location × year) was conducted within the INVITE project (https://www.h2020-invite.eu/) between 2021 and 2023 in 18 European locations (8 countries) spread along a climatic transect for temperature and evaporative demand in both rain-fed and irrigated conditions. The maize panel used in this dataset consisted of 31 highly successful commercial maize hybrids released on the European market from 2009 to 2020. The panel was designed to show a limited range of maturity class, from mid-early to mid-late covering the largest growing area in Europe. It was genotyped with a 600K Axiom Affymetrix array. To account for genetic relatedness among individuals, we computed the kinship matrix from the marker dataset using the VanRaden method implemented in the ASRgenomics package in R. This method estimates the realized additive genomic relationship matrix based on marker allele frequencies, providing a robust measure of pairwise genetic similarity. The resulting kinship matrix was incorporated into subsequent analyses. 

\textbf{Marker dataset 2}

This dataset was also generated in the INVITE project. It comprises of 80 European registered maize hybrid varieties that were tested in most European value for cultivation and use (VCU) trials between 2003-2018. The varieties were genotyped with a 600K Axiom Affymetrix array and kinship matrix was computed using the VanRaden method. 

\textbf{Marker dataset 3}

Within the Synbreed project (\citealt{alb}), a total of 1,380 doubled haploid maize lines were generated and genotypes with 1,152 SNP markers randomly distributed across the genome using the Illumina VeraCode technology. A subset of 698 testcrosses made from these doubled haploid lines were considered in \cite{est1}. For these, the kinship matrix is available as computed in VanRaden form.

\section{The Linear Mixed Model}

In this chapter we specify the LMM that we use for the design purposes (Section~\ref{3.1}), and we derive the formulas for the mean squared error (MSE) matrices for the BLUPs of the genotype effects and their pairwise linear contrasts (Section~\ref{3.2}).

\subsection{Model Specification}\label{3.1}

We use a linear mixed model for field plot data for replicated multi-environment trials conducted in multiple locations within each sub-region across multiple years. A salient feature of the model is that the genetic effects are allowed to be correlated according to the kinship among the tested genotypes as modeled via the kinship matrix. In addition, the model entails a genetic correlation between the sub-regions, thus permitting information to be shared between sub-regions. The novelty compared to earlier work (\citealt{pru9}, \citealt{pru10}) is the inclusion of the kinship matrix. 

In the following, we assume that field trials in each environment are laid out according to a randomized complete block design (RCBD).
For our purposes we use the LMM in which the yield of genotype $k$ in block $l$ in location $j$ in sub-region $i$ in year $h$ is given by
\begin{equation}\label{mod}
Y_{ijhkl}=\mu_i+\eta_h+\lambda_{ij}+\beta_{ih}+\delta_{ijh}+b_{ijhl}+\alpha_{ik}+\omega_{hk}+\gamma_{ijk}+\tau_{ihk}+\phi_{ijhk}+\varepsilon_{ijhkl}
\end{equation}
for $i=1 \dots, P$, $j=1, \dots, J_i$, $h=1, \dots, H$, $k=1, \dots, K$, $l=1, \dots, L$, where 

$\mu_i$ denotes the mean (fixed) effect of the $i$-th sub-region, 

$\eta_h$ is the effect of the $h$-th year, 

$\lambda_{ij}$ is the effect of the $j$-th location within the $i$-th sub-region, 

$\beta_{ih}$ is the effect of the $i$-th sub-region in the $h$-th year, 

$\delta_{ijh}$ is the effect of location $j$ within sub-region $i$ in the $h$-th year,

$b_{ijhl}$ is the effect of the $l$-th replication (block) in location $j$ in sub-region $i$ in the $h$-th year, 

$\alpha_{ik}$ is the effect of genotype $k$ in sub-region $i$,

 $\omega_{hk}$ is the effect of genotype $k$ in the $h$-th year,

 $\gamma_{ijk}$ denotes the effect of the $k$-th genotype in the $j$-th location within the $i$-th sub-region,

 $\tau_{ihk}$ is the effect of genotype $k$ within sub-region $i$ in the $h$-th year, 

$\phi_{ijhk}$ is the effect of genotype $k$ in location $j$ in the $h$-th year, 

and $\varepsilon_{ijhkl}$ denotes the observational error. 

All effects besides $\mu_i$ are random. The observational errors and all random effects, besides the genotype effects for the same genotype, are assumed to be uncorrelated and to have zero mean. The variances are given by 
$\mathrm{var}(\varepsilon_{ijkl})=\sigma^2$, 
$\mathrm{var}(\eta_h)=\sigma^2_{\eta}$, 
$\mathrm{var}(\lambda_{ij})=\sigma^2_{\lambda}$, 
$\mathrm{var}(\beta_{ih})=\sigma^2_{\beta}$,
$\mathrm{var}(\delta_{ijh})=\sigma^2_{\delta}$,
$\mathrm{var}(b_{ijhl})=\sigma^2_{b}$, 
$\mathrm{var}(\omega_{hk})=\sigma^2_{\omega}$,
$\mathrm{var}(\gamma_{ijk})=\sigma^2_{\gamma}$, 
$\mathrm{var}(\tau_{ihk})=\sigma^2_{\tau}$, 
and $\mathrm{var}(\phi_{ijhk})=\sigma^2_{\phi}$.
The vector of all genotype effects for all genotypes is given by $\mbox{\boldmath{$\alpha$}}=(\mbox{\boldmath{$\alpha$}}_{1}^\top, \dots, \mbox{\boldmath{$\alpha$}}_{K}^\top)^\top$, where $\mbox{\boldmath{$\alpha$}}_k=(\alpha_{1k}, \dots, \alpha_{Pk})^\top$ is the vector of all genotype$\times$sub-region effects for genotype $k$. The covariance matrix of the genotype effects is given by $\mathrm{Cov}(\mbox{\boldmath{$\alpha$}})=\mathbf{U}$, where $\mathbf{U}$ is some positive definite matrix. 

In multivariate analysis the Kronecker structure of the covariance matrix is typically used (e.\,g. \citealt{gil}). For the present model the covariance matrix of the genotype effects may be written as $\mathbf{U}=\mathbf{N}\otimes\mathbf{V}$, where $\mathbf{N}$ and $\mathbf{V}$ are positive definite and $\otimes$ denotes the Kronecker product. In this case the covariance matrix of the genotype$\times$sub-region effects for genotype $k$ is given by $\mathrm{Cov}(\mbox{\boldmath{$\alpha$}}_k)=n_{kk}\mathbf{V}$, where $n_{kk}$ denotes the $k$-th diagonal element of matrix $\mathbf{N}$, and $\mathbf{N}$ denotes the genotypes relationship matrix (or kinship matrix).

Note that model \eqref{mod} differs from the LMM considered in \cite{pru10} only by the covariance matrix of the genotype effects. For the particular case of uncorrelated genotype effects with the same covariance matrix $\mathbf{V}$ for each genotype, the covariance matrix for all genotypes simplifies to $\mathbf{U}=\mathbb{I}_K\otimes\mathbf{V}$, where $\mathbb{I}_K$ is the $K\times K$ identity matrix. The LMM with correlated genotype effects has been considered for one-year experiments in \cite{prus2025computing}.

The present model in its general form describes the cross-classification case: experiments with the same locations each year. The cross-classified case assumes that the exact same locations are used in every year. If locations change each year, the factor location is nested within years (\citealt{patt}). Note that for annual crops, rotation requirements mean that the trials will never be planted in the exact same place each year, even if the location names are the same. This leads some authors to generally prefer the nested model (\citealt{patt}, \citealt{for}). The nested model can be derived from the cross-classified model by dropping out effects indexed by location but not by year, and absorbing them into the corresponding effect indexed by both location and year. This leads to the following particular case of model \eqref{mod}:
\begin{equation}\label{mod2}
Y_{ijhkl}=\mu_i+\eta_h+\beta_{ih}+\delta_{ijh}+b_{ijhl}+\alpha_{ik}+\omega_{hk}+\tau_{ihk}+\phi_{ijhk}+\varepsilon_{ijhkl}.
\end{equation}
This model differs from the previous one by the missing location within sub-region effects $\lambda_{ij}$ and genotype$\times$location effects $\gamma_{ijk}$. Note that variances, $\mathrm{var}(\delta_{ijh})$ and $\mathrm{var}(\phi_{ijhk})$, of the location$\times$year effects and the genotype$\times$location$\times$year effects in model \eqref{mod2}, coincide with $\sigma^2_{\lambda}+\sigma^2_{\delta}$ and $\sigma^2_{\gamma}+\sigma^2_{\phi}$, respectively, in model \eqref{mod}. We will consider both models \eqref{mod} and \eqref{mod2}.

In this work our main focus is on both the genotype effects $\mbox{\boldmath{$\alpha$}}_1, \dots, \mbox{\boldmath{$\alpha$}}_K$ and their pairwise linear contrasts
\begin{equation*}
\mbox{\boldmath{$\theta$}}^{k,k'}=\mbox{\boldmath{$\alpha$}}_k-\mbox{\boldmath{$\alpha$}}_{k'},\, k, k'=1, \dots, K,\, k\neq k'.
\end{equation*}
We search for the numbers of locations $J_1, \dots, J_P$ that are optimal for the prediction (BLUP) of the genotype effects themselves or their pairwise linear contrasts.

\subsection{The MSE matrices for the BLUPs}\label{3.2}

The performance of the prediction may be measured by its MSE matrix. For the BLUP, the MSE matrix can be derived based on Henderson's approach. The next lemma provides the MSE matrix of the BLUP for the genotype effects in model \eqref{mod}.
\begin{lem}\label{lem1}
For model \eqref{mod}, the MSE matrix of the BLUP $\hat{\mbox{\boldmath{$\alpha$}}}$ of the genotype effects $\mbox{\boldmath{$\alpha$}}$ is given by 
\begin{equation}\label{mse}
\mathrm{Cov}(\hat{\mbox{\boldmath{$\alpha$}}}-\mbox{\boldmath{$\alpha$}})=\left\{\frac{1}{c}\mathbf{T}\otimes[(\mathbf{F}^\top\mathbf{F})^{-1}+\mathbf{R}]^{-1}+\mathbf{U}^{-1}\right\}^{-1},
\end{equation}
where
\begin{equation*}
\mathbf{F}=\textrm{block-diag}(\mathds{1}_{J_1}, \dots, \mathds{1}_{J_P}),
\end{equation*}
\begin{equation*}
\mathbf{R}=\frac{\sigma_\tau^2}{cH}\mathbb{I}_P+\frac{\sigma_\omega^2}{cH}\mathds{1}_P\mathds{1}_P^\top,
\end{equation*}
\begin{equation*}
c=\sigma_\gamma^2+\frac{1}{H}\left(\sigma_\phi^2+\frac{1}{L}\sigma^2\right),
\end{equation*}
$\mathbf{T}=\mathbb{I}_K-\frac{1}{K}\mathds{1}_K\mathds{1}_K^\top$, $\mathds{1}_n$ denotes the vector of length $n$ with all entries equal to $1$, and $\textrm{block-diag}(\mathds{1}_{J_1}, \dots, \mathds{1}_{J_P})$ is the block diagonal matrix with the blocks $\mathds{1}_{J_1}, \dots, \mathds{1}_{J_P}$.
\end{lem}
The proof of Lemma~\ref{lem1} is deferred to Web Appendix~\ref{pl1}.

The next Corollary provides the MSE matrix of the BLUP of the pairwise linear contrasts. The vector $\mbox{\boldmath{$\theta$}}$ of pairwise linear contrasts for all $n=\frac{K(K-1)}{2}$ pairs of genotypes is given by $\mbox{\boldmath{$\theta$}}=(\mbox{\boldmath{$\theta$}}_1^\top, \dots, \mbox{\boldmath{$\theta$}}_{K-1}^\top)^\top$, where $\mbox{\boldmath{$\theta$}}_s=(\mbox{\boldmath{$\theta$}}_{s,s+1}^\top, \dots, \mbox{\boldmath{$\theta$}}_{s,K}^\top)^\top$ for $s=1,\dots,K-1$. 
The contrasts may also be written as 
\begin{equation*}
    \mbox{\boldmath{$\theta$}}_{k,k'}=((\mathbf{e}_k-\mathbf{e}_{k'})^\top\otimes\mathbb{I}_P)\,\mbox{\boldmath{$\alpha$}},
\end{equation*}
and then the vectors for groups of contrasts $\mbox{\boldmath{$\theta$}}_s$ can be written as 
\begin{equation*}
    \mbox{\boldmath{$\theta$}}_s=(\mathcal{C}_s\otimes \mathbb{I}_P)\,\mbox{\boldmath{$\alpha$}},
\end{equation*} 
where $\mathcal{C}_s$ is the $(K-s)\times K$ matrix given by
\begin{equation*}
\mathcal{C}_s=(0_{(K-s)\times(s-1)},\, \mathds{1}_{K-s},\, -\mathbb{I}_{K-s}),
\end{equation*} 
for $s=2, \dots, K-1$, and 
\begin{equation*}
\mathcal{C}_1=(\mathds{1}_{K-1},\, -\mathbb{I}_{K-1}).
\end{equation*}

\begin{fol}\label{fol}
For model \eqref{mod}, the MSE matrix of the BLUP $\hat{\mbox{\boldmath{$\theta$}}}$ of the pairwise linear contrasts $\mbox{\boldmath{$\theta$}}$ is given by 
\begin{equation}\label{mse2}
\mathrm{Cov}(\hat{\mbox{\boldmath{$\theta$}}}-\mbox{\boldmath{$\theta$}})=\left(\mathcal{C}\otimes\mathbb{I}_P\right)\left\{\frac{1}{c}\mathbf{T}\otimes[(\mathbf{F}^\top\mathbf{F})^{-1}+\mathbf{R}]^{-1}+\mathbf{U}^{-1}\right\}^{-1}\left(\mathcal{C}^\top\otimes\mathbb{I}_P\right),
\end{equation}
where $\mathcal{C}$ is the $(n\times K)$ contrasts matrix given by
\begin{equation*}
\mathcal{C}=(\mathcal{C}_1^\top, \dots, \mathcal{C}_{K-1}^\top).
\end{equation*}
\end{fol} 

\begin{bem}
For model \eqref{mod2}, the MSE matrices of the BLUP $\hat{\mbox{\boldmath{$\alpha$}}}$ of the genotype effects $\mbox{\boldmath{$\alpha$}}$ and of the BLUP $\hat{\mbox{\boldmath{$\theta$}}}$ of the pairwise linear contrasts $\mbox{\boldmath{$\theta$}}$ can be obtained by replacing, respectively, in formulas \eqref{mse} and \eqref{mse2} the constant $c$ with $\tilde{c}=\frac{1}{H}\left(\sigma_\phi^2+\frac{1}{L}\sigma^2\right)$.
\end{bem}

\section{Optimal Design}

In this chapter we present analytical results for optimal allocations of trials, i.\,e. for optimal designs. We derive the results for model \eqref{mod} only, since model~\eqref{mod2} is the particular case of model \eqref{mod} with $\sigma^2_{\lambda}=\sigma^2_{\gamma}=0$. In Section~\ref{4.1} we present the standard and the weighted A-criteria for the genotype effects and their pairwise linear contrasts. Section~\ref{part cases} considers some practically relevant particular cases of the covariance structure for the genotype effects. Section~\ref{cm} is related to the computation of optimal and highly efficient designs. 

We define the (exact) designs as follows:
\begin{equation*} \xi:=\left( \begin{array}{ccc} J_1 & ... & J_P
\end{array}\right),
\end{equation*}
where $J_1, \dots, J_P$ are the numbers of locations in sub-regions $1, \dots P$, respectively.
To make use of convex optimization theory later, we use approximate designs, defined as 
\begin{equation*}\xi:=\left( \begin{array}{ccc} w_1 & ... & w_P
\end{array}\right),
\end{equation*}
where $w_1, ..., w_P$ denote the weights of locations in sub-regions 
that satisfy only the conditions $\sum_{i=1}^P{w_i}=1$ and $w_i\geq 0$. Note that the numbers of locations $J_i=w_iJ$ resulting from approximate designs are not necessarily integer, i.\,e. optimal approximate designs are not directly applicable.

Further we use the following notation for the  moment matrix, which includes all the information about designs:
\begin{equation*}
\mathbf{M}(\xi)=\textrm{diag}(w_1, \dots, w_P).
\end{equation*} 
For exact designs it holds that
$\mathbf{M}(\xi)=\frac{1}{J}\mathbf{F}^\top\mathbf{F}$.

Then we can extend the definition of MSE matrices \eqref{mse} and \eqref{mse2} for approximate designs and rewrite them (neglecting the constant multiplicator $\frac{c}{J}$), respectively, as
\begin{equation}\label{mse1}
\mathrm{MSE}_{\alpha}(\xi)=\left\{\mathbf{T}\otimes[\mathbf{M}(\xi)^{-1}+\tilde{\mathbf{R}}]^{-1}+\tilde{\mathbf{U}}^{-1}\right\}^{-1},
\end{equation}
where $\tilde{\mathbf{R}}=J\mathbf{R}$ and $\tilde{\mathbf{U}}=\frac{J}{c}\mathbf{U}$, and
\begin{equation}\label{mse3}
\mathrm{MSE}_{\theta}(\xi)=\left(\mathcal{C}\otimes\mathbb{I}_P\right)\left\{\mathbf{T}\otimes[\mathbf{M}(\xi)^{-1}+\tilde{\mathbf{R}}]^{-1}+\tilde{\mathbf{U}}^{-1}\right\}^{-1}\left(\mathcal{C}^\top\otimes\mathbb{I}_P\right).
\end{equation}

\subsection{Design Criteria}\label{4.1}

For models with genetic correlation between genotypes, the weighted \textit{A}-criterion for prediction of the genotype effects is defined in \cite{prus2025computing} as
\begin{equation}\label{wacr11}
\Phi_{A_w, \alpha}=\sum_{k=1}^K\sum_{i=1}^P\ell_i\mathrm{var}(\hat{\alpha}_{ik}-\alpha_{ik}),
\end{equation}
where  $\ell_1, \dots, \ell_P$ denote some sub-regional coefficients. These coefficients could be chosen in proportion to the arable area of each sub-region or to the size of the population in each sub-region.
This criterion can also be written as
\begin{equation}\label{wacr1}
\Phi_{A_w, \alpha}=\mathrm{tr}\left(\mathrm{Cov}(\hat{\mbox{\boldmath{$\alpha$}}}-\mbox{\boldmath{$\alpha$}})\left(\mathbb{I}_K\otimes\mathcal{L}\right)\right),
\end{equation}
where $\mathcal{L}=\textrm{diag}(\ell_1, \dots, \ell_P)$, which is the $P\times P$ diagonal matrix with the diagonal elements $\ell_1, \dots, \ell_P$. 

For approximate designs, the definition of the weighted \textit{A}-criterion is extended by replacing $\mathrm{Cov}(\hat{\mbox{\boldmath{$\alpha$}}}-\mbox{\boldmath{$\alpha$}})$ with $\mathrm{MSE}_{\alpha}(\xi)$ in formula \eqref{wacr1}. The next theorem provides the explicit form of the criterion, neglecting constants that have no influence on designs.

\begin{thm}\label{t3}
The weighted \textit{A}-criterion for the prediction of the genotype effects is given for approximate designs by
\begin{equation}\label{wacrappr}
\Phi_{A_w, \alpha}(\xi)=\mathrm{tr}\left[\left( \mathbb{I}_K\otimes\mathbf{M}(\xi)+\mathbf{B}\right)^{-1}\mathbf{H}\right],
\end{equation}
where  
\begin{equation*}
    \mathbf{B}=\left[\mathbb{I}_K\otimes \tilde{\mathbf{R}}+(\mathbf{T}\otimes\mathbb{I}_P)\,\tilde{\mathbf{U}}\,(\mathbf{T}\otimes\mathbb{I}_P)\right]^{-1}
\end{equation*} and 
\begin{equation*}\mathbf{H}=\mathbf{B}\,(\mathbf{T}\otimes\mathbb{I}_P)\,\tilde{\mathbf{U}}(\mathbb{I}_K\otimes\mathcal{L})\tilde{\mathbf{U}}(\mathbf{T}\otimes\mathbb{I}_P)\,\mathbf{B}.
\end{equation*}
\end{thm}
The proof of Theorem~\ref{t3} is deferred to Web Appendix~\ref{pt1}.

Note that $\mathbf{N}=\mathbb{I}_K$, weighted \textit{A}-criterion \eqref{wacrappr} coincides with the weighted \textit{A}-criterion in \cite{pru10}.

The standard \textit{A}-criterion is defined as
\begin{equation}\label{acr1}
\Phi_{A,\alpha}=\mathrm{tr}\left(\mathrm{Cov}(\hat{\mbox{\boldmath{$\alpha$}}}-\mbox{\boldmath{$\alpha$}})\right),
\end{equation}
which can be recognized as the particular case of the weighted \textit{A}-criterion with $\mathcal{L}=\mathbb{I}_P$.
\begin{fol}\label{f1}
The \textit{A}-criterion for the prediction of the genotype effects is given for approximate designs by
\begin{equation}\label{acrappr}
\Phi_{A,\alpha}(\xi)=\mathrm{tr}\left[\left( \mathbb{I}_K\otimes\mathbf{M}(\xi)+\mathbf{B}\right)^{-1}\mathbf{H}_1\right],
\end{equation}
where $\mathbf{H}_1=\mathbf{B}\,(\mathbf{T}\otimes\mathbb{I}_P)\,\tilde{\mathbf{U}}^2(\mathbf{T}\otimes\mathbb{I}_P)\,\mathbf{B}$.
\end{fol}

We define the weighted \textit{A}-criterion for the prediction of the pairwise linear contrasts in model \eqref{mod} for exact designs as
\begin{equation}\label{wacr2}
\Phi_{A_w,\theta}=\sum_{k=1}^{K-1}\sum_{k'=k+1}^K\left[\mathrm{Cov}(\hat{\theta}_{k,k'}-\theta_{k,k'})\mathcal{L}\right],
\end{equation}
where $\mathrm{Cov}(\hat{\theta}_{k,k'}-\theta_{k,k'})$ are the diagonal blocks of $\mathrm{MSE}_{\theta}$ with size $P\times P$.

Alternatively, criterion \eqref{wacr2} can be written as
\begin{equation}\label{wacrdef}
\Phi_{A_w, \theta}=\mathrm{tr}\left[\mathrm{Cov}(\hat{\theta}-\theta)\left(\mathbb{I}_n\otimes\mathcal{L}\right)\right].
\end{equation}

For approximate designs, we replace $\mathrm{Cov}(\hat{\theta}-\theta)$ with  
    $\mathrm{MSE}_{\theta}(\xi)$, and we obtain (neglecting constants) the following result.

\begin{fol}\label{t4}
The weighted \textit{A}-criterion for the prediction of the pairwise linear contrasts is given for approximate designs by
\begin{equation}\label{wacrapprdef}
\Phi_{A_w, \theta}(\xi)=\mathrm{tr}\left[\left( \mathbb{I}_K\otimes\mathbf{M}(\xi)+\mathbf{B}\right)^{-1}\tilde{\mathbf{H}}\right],
\end{equation}
where
\begin{equation*}\tilde{\mathbf{H}}=\mathbf{B}\,(\mathbf{T}\otimes\mathbb{I}_P)\,\tilde{\mathbf{U}}(\mathbf{T}\otimes\mathcal{L})\tilde{\mathbf{U}}(\mathbf{T}\otimes\mathbb{I}_P)\,\mathbf{B}.
\end{equation*}
\end{fol}

For the proof of Corollary~\ref{t4} see Web Appendix~\ref{pc3}.

After replacing $\mathcal{L}$ by $\mathbb{I}_P$ in formula \eqref{wacrapprdef}, we obtain the standard \textit{A}-criterion for the prediction of the contrasts.
\begin{fol}\label{t5}
The \textit{A}-criterion for the prediction of the pairwise linear contrasts is given for approximate designs by
\begin{equation}\label{acrapprdef}
\Phi_{A, \theta}(\xi)=\mathrm{tr}\left[\left( \mathbb{I}_K\otimes\mathbf{M}(\xi)+\mathbf{B}\right)^{-1}\tilde{\mathbf{H}}_1\right],
\end{equation}
where
\begin{equation*}\tilde{\mathbf{H}}_1=\mathbf{B}\,(\mathbf{T}\otimes\mathbb{I}_P)\,\tilde{\mathbf{U}}(\mathbf{T}\otimes\mathbb{I}_P)\tilde{\mathbf{U}}(\mathbf{T}\otimes\mathbb{I}_P)\,\mathbf{B}.
\end{equation*}
\end{fol}

\begin{bem}
For model \eqref{mod2}, the standard and the weighted A-criteria for the prediction of the genotype effects or their pairwise linear contrasts  can be obtained by replacing the constant $c$ with $\tilde{c}=\frac{1}{H}\left(\sigma_\phi^2+\frac{1}{L}\sigma^2\right)$ in matrices $\tilde{\mathbf{R}}$ and $\tilde{\mathbf{U}}$ in formulas \eqref{acrappr} and \eqref{wacrappr} or \eqref{acrapprdef} and \eqref{wacrapprdef}, respectively.
\end{bem}

\subsection{Particular Cases}\label{part cases}

In this section we assume the Kronecker product structure $\mathbf{U}=\mathbf{N}\otimes\mathbf{V}$ of the covariance matrix of the genotype effects. In particular, we consider the compound symmetry (SC) structure and the block-diagonal structure with CS blocks of the kinship matrix $\mathbf{N}$.

\begin{thm}\label{CS1}
Let the kinship matrix $N$ have the compound symmetry structure $\mathbf{N}=a_1\mathbb{I}_K+a\mathds{1}_K\mathds{1}_K^\top$, where $a=\sigma^2_\alpha\,r$ and  $a_1=\sigma^2_\alpha(1-r)$ with $\sigma_\alpha^2>0$ and $r\in[0,1)$. Then, optimal designs for the prediction of the genotype effects coincide with optimal designs for their pairwise linear contrasts, and are independent of the number of genotypes $K$ and parameter $a$. Moreover, optimal designs for the Bayesian linear criterion 
\begin{equation}\label{bacr}
\Phi^B_{A}(\xi)=\mathrm{tr}\left\{\left[\mathbf{M}(\xi)+\mathbf{S}^{-1}\right]^{-1}\mathbf{S}^{-1}\tilde{\mathbf{V}}^2\mathbf{S}^{-1}\right\},
\end{equation}
where $\mathbf{S}=\tilde{\mathbf{R}}+a_1\tilde{\mathbf{V}}$ and $\tilde{\mathbf{V}}=\frac{J}{c}\mathbf{V}$, are optimal for the standard \textit{A}-criteria \eqref{acrappr} and \eqref{acrapprdef},
and the optimal designs for the Bayesian linear criterion
\begin{equation}\label{bwacr}
\Phi^B_{A_w}(\xi)=\mathrm{tr}\left\{\left[\mathbf{M}(\xi)+\mathbf{S}^{-1}\right]^{-1}\mathbf{S}^{-1}\tilde{\mathbf{V}}\mathcal{L}\tilde{\mathbf{V}}\mathbf{S}^{-1}\right\}
\end{equation}
are optimal for the weighted \textit{A}-criteria \eqref{wacrappr} and \eqref{wacrapprdef}. 
\end{thm}
The proof of Theorem~\ref{CS1} is deferred to Web Appendix~\ref{pt2}.

Note that for $a_1=1$ optimal designs for the standard and the weighted \textit{A}-criteria coincide with the optimal designs obtained in \cite{pru10}.

Further we consider the particular case of $f$ families of genotypes with $m$ genotypes in each family, i.\,e. $K=f\,m$, and the block-diagonal structure of the kinship matrix with CS blocks.

\begin{thm}\label{CS2}
Let the kinship matrix $N$ have the following block-diagonal structure with compound symmetry blocks: $\mathbf{N}=\mathbb{I}_f\otimes\left(b_1\mathbb{I}_m+b\mathds{1}_m\mathds{1}_m^\top\right)$, where $b=\sigma^2_\alpha\,r$ and  $b_1=\sigma^2_\alpha(1-r)$ with $\sigma_\alpha^2>0$ and $r\in[0,1)$. Then, optimal designs for the prediction of the genotype effects coincide with optimal designs for their pairwise linear contrasts. Moreover, optimal designs for the CBRC criterion 
\begin{equation}\label{cbrc1}
\Phi^{CBRC}_{A}(\xi)=\mathrm{tr}\left\{\left[\mathbf{M}(\xi)+\mathbf{S}_1^{-1}\right]^{-1}\mathbf{Q}_1\right\}+\mathrm{tr}\left\{\left[\mathbf{M}(\xi)+\mathbf{S}_2^{-1}\right]^{-1}\mathbf{Q}_2\right\},
\end{equation}
where $\mathbf{Q}_1=f(m-1)\mathbf{S}_1^{-1}\mathbf{V}_1^2\mathbf{S}_1^{-1}$, $\mathbf{Q}_2=(f-1)\mathbf{S}_2^{-1}\mathbf{V}_2^2\mathbf{S}_2^{-1}$, $\mathbf{S}_i=\tilde{\mathbf{R}}+\mathbf{V}_i$, for $i=1,2$, $\mathbf{V}_1=b_1\tilde{\mathbf{V}}$, and $\mathbf{V}_2=(mb+b_1)\tilde{\mathbf{V}}$,
are optimal for the standard \textit{A}-criteria \eqref{acrappr} and \eqref{acrapprdef},
and optimal designs for the CBRC criterion
\begin{equation}\label{cbrc2}
\Phi^{CBRC}_{A_w}(\xi)=\mathrm{tr}\left\{\left[\mathbf{M}(\xi)+\mathbf{S}_1^{-1}\right]^{-1}\mathbf{Q}_3\right\}+\mathrm{tr}\left\{\left[\mathbf{M}(\xi)+\mathbf{S}_2^{-1}\right]^{-1}\mathbf{Q}_4\right\},
\end{equation}
where $\mathbf{Q}_3=f(m-1)\mathbf{S}_1^{-1}\mathbf{V}_1\mathcal{L}\mathbf{V}_1\mathbf{S}_1^{-1}$ and $\mathbf{Q}_4=(f-1)\mathbf{S}_2^{-1}\mathbf{V}_2\mathcal{L}\mathbf{V}_2\mathbf{S}_2^{-1}$,
are optimal for the weighted \textit{A}-criteria \eqref{wacrappr} and \eqref{wacrapprdef}. 
\end{thm}
The proof of Theorem~\ref{CS2} is deferred to Web Appendix~\ref{pt3}.

Note that optimal designs for \eqref{cbrc1} and \eqref{cbrc2} depend on the number of families $f$ and the number of genotypes per family $m$ as well as on the variance $\sigma_\alpha^2$ and the covariance parameter $r$. Moreover, the designs depend on the number of genotypes $K$, since $K=f\,m$. This dependence will be illustrated by an example in Section~\ref{rde}. 

Note also that in the trivial case of only one family of genotypes, i.\,e. $f=1$, criteria \eqref{cbrc1} and \eqref{cbrc2} simplify to \eqref{bacr} and \eqref{bwacr}, respectively. The other trivial case with only one genotype per family, i.\,e. $m=1$, coincides with the case of uncorrelated genotype effects for different genotypes that has been discussed in \cite{pru10}. The latter also concerns the case of zero covariance, i.\,e. $r=0$.

For $f>1$ and $m>1$, criteria \eqref{cbrc1} and \eqref{cbrc2} can be rewritten in the form of K-Bayesian linear criteria. The next corollary provides the result.

\begin{fol}
    The CBRC criteria \eqref{cbrc1} and \eqref{cbrc2} coincide with the K-Bayesian linear criteria 
    \begin{equation}\label{kb1}
        \Phi^{KB}_{A}(\xi)=\mathrm{tr}\left[\left( \mathbb{I}_2\otimes\mathbf{M}(\xi)+\tilde{\mathbf{S}}^{-1}\right)^{-1}\mathbf{Q}\right],
    \end{equation}
    where $\tilde{\mathbf{S}}=\text{block-diag}(\mathbf{S}_1,\mathbf{S}_2)$, $\mathbf{Q}=\text{block-diag}(\mathbf{Q}_1,\mathbf{Q}_2)$,
    and
        \begin{equation}\label{kb2}
        \Phi^{KB}_{A_w}(\xi)=\mathrm{tr}\left[\left( \mathbb{I}_2\otimes\mathbf{M}(\xi)+\tilde{\mathbf{S}}^{-1}\right)^{-1}\mathbf{Q}'\right],
    \end{equation}
    where $\mathbf{Q}'=\text{block-diag}(\mathbf{Q}_3,\mathbf{Q}_4)$,
    respectively.
\end{fol}
Note that the dimension of the information matrix $\mathbb{I}_2\otimes\mathbf{M}(\xi)+\tilde{\mathbf{S}}^{-1}$ in the obtained K-Bayesian criteria \eqref{kb1} and \eqref{kb2} is only $2\,P$, which is much lower than the dimension $KP$ of the information matrix $\mathbb{I}_K\otimes\mathbf{M}(\xi)+\mathbf{B}$ in the original standard and weighted \textit{A}-criteria \eqref{acrappr} and \eqref{wacrappr}. Hence, the computation of optimal allocations of trials becomes considerably less complicated after using Theorem~\ref{CS2}.

\begin{bem}
The Bayesian criteria \eqref{bacr} and \eqref{bwacr}, the CBRC criteria \eqref{cbrc1} and \eqref{cbrc2},
and the K-Bayesian criteria \eqref{kb1} and \eqref{kb2} can be adjusted for model \eqref{mod2} by replacing the constant $c$ with $\tilde{c}=\frac{1}{H}\left(\sigma_\phi^2+\frac{1}{L}\sigma^2\right)$ in matrices $\tilde{\mathbf{R}}$ and $\tilde{\mathbf{V}}$.
\end{bem}

\subsection{Computation of Optimal and Highly Efficient Designs}\label{cm}

Most algorithms for computing optimal designs are tailored to the standard linear regression model, approximate designs, and the most common optimality criteria, such as D- and A-optimality (see, e.g., \cite{mandal2015algorithmic} for an overview).

When the problem size is rather small, mathematical programming techniques, most notably mixed-integer second-order cone programming (MISOCP), can be applied to compute exact optimal designs (\cite{sag}). For models similar to ours, MISOCP has been successfully used to compute designs for a very limited number of genotypes \cite{prus2025computing}. This approach is effective for small problems, but becomes computationally demanding for larger design spaces, such as the one considered here, thus, we have to resort to heuristic methods. The most popular heuristic procedure for computing exact designs in linear models is the KL-exchange algorithm (\cite{atk1}).

However, in this paper, we adopt a different approach from KL-exchange, primarily due to the complicated form of the criterion and to be able to incorporate linear constraints into the optimization process. Specifically, we employ a modified version of the algorithm proposed in \cite{rad}, originally developed for exact D-optimal designs with linear resource constraints. The underlying philosophy of this algorithm allows its application to a broad class of problems, as long as the optimality criterion remains convex. This method has previously been used successfully for computing exact designs in multiple-group mixed models with a structure similar to ours in \cite{fil}.

\section{Real Data Example 
}\label{rde}

We use the real data example for maize in India from \cite{kle}, that considers a three-year experiment with five sub-regions, i.\,e. $H=3$ and $P=5$. The total number of locations $J$ is not fixed in the experiment. Therefore, we compute optimal designs for different values of the total number of locations. The estimated values of the variance parameters needed for the computation of optimal designs in model \eqref{mod} are given by 
$\sigma_\omega^2=31$, $\sigma_\tau^2=18$, $\sigma_\gamma^2=160$ and $\sigma_\phi^2+\frac{1}{L}\sigma^2=333$ (for more details see Table~1 in \citealt{pru10}). For model \eqref{mod2} we have the same values of $\sigma_\omega^2$ and $\sigma_\tau^2$, and $\sigma_\phi^2+\frac{1}{L}\sigma^2=493$ (Table~2 in \citealt{pru10}). The variance-covariance matrix of the genotype effects for the $k$-th genotype is $n_{kk}\mathbf{V}$, where $\mathbf{V}$ is given by
\begin{equation*}
\mathbf{V}=\left(\begin{array}{ccccc}
	567 & 254 & 239 & 485 & 328 \\ 254 & 155 & 118 & 240 & 162 \\ 239 & 118 & 155 & 226 & 153 \\ 485 & 240 & 226 & 488 & 310 \\ 328 & 162 & 153 & 310 & 215
\end{array}\right).
\end{equation*}

For the weighted \textit{A}-criterion we use sub-regional coefficients related to the areas of the sub-regions
\begin{equation*}
\begin{array}{ccccc} \ell_1 & \ell_2 & \ell_3 & \ell_4 & \ell_5 \\ 813685 & 432716 & 477365 & 995298 & 1174818
\end{array}.
\end{equation*}

For the present experiment it is necessary that there is at least one location in each sub-region, i.\,e. $J_i\geq 1$,\, $i=1, ..., 5$. Therefore, we deal with constrained designs, that can be however easily handled by the computational algorithm described in Section~\ref{cm}.

\subsection{Optimal and Highly Efficient Designs for Three Kinship Matrices}\label{3 kinship matrices}

We consider optimal and highly efficient exact designs for the three kinship matrices described in Section~\ref{data} with $K=31$, $K=80$, and $K=698$ genotypes, and for various values of the total number of locations $J$. 

Table~\ref{T:3k} depicts the exact designs ($J_1, \dots, J_5$) obtained for the prediction of the genotype effects (blocks 1,3,5,7) and for their pairwise linear contrasts (2,4,6,8), for the standard (blocks 1,2,5,6) and the weighted (blocks 3,4,7,8) \textit{A}-criterion, in the two models \eqref{mod} (blocks 1-4) and \eqref{mod2} (blocks 5-8). 



The MSE related values denoted by $\mathrm{MSE}_{Tr}$ in the table coincide for the genotype effects or their pairwise linear contrasts with the trace of MSE matrix \eqref{mse1} or \eqref{mse3}, respectively. 

As we can see in the table, for both models the obtained designs for the standard \textit{A}-criterion assign the highest numbers of locations to the sub-regions with the highest variance (largest diagonal elements of matrix $\mathbf{V}$), i.\,e. sub-regions "1" and "4". For the weighted criterion, the designs assign more locations to the sub-regions with larger values of the sub-regional coefficients (sub-regions "5", "4", and "1"). Results for model~\eqref{mod} and model~\eqref{mod2} are in general similar to each other. However, for model~\eqref{mod2} the designs turn out to be a bit more balanced than for model~\eqref{mod}.  Note that a similar behavior of optimal designs has been observed in \cite{pru10}. 

In the table, we do not observe any essential difference between the designs for the prediction of genotype effects and their pairwise linear contrasts. However, the designs are in general not the same. 

As we can also see in Table~\ref{T:3k}, in cases with large number of genotypes ($K=698$) optimal designs tend to balanced designs. This behavior of designs is true for the present particular kinship matrices. However, it does not hold in general, as it will be illustrated in Section~\ref{CS block-diag}.

\subsection{Designs with Additional Cost and Resource Constraints}

Besides the total number of locations $J$ and the restriction of at least one location in each sub-region, the computational method described in Section~\ref{cm} makes it possible to introduce other reasonable constraints. To illustrate this, we have chosen the case of 80 genotypes, and we focused on the standard \textit{A}-criterion for the prediction of genotype effects in model \eqref{mod}.

We will introduce the following sets of additional constraints to the problem:

\begin{itemize}
  \item $C_1$: the minimum number of locations in each sub-region is 2
  \item $C_2$: the minimum number of locations in each sub-region is 2 and the maximum number of locations in each sub-region is $J/4$
  \item $C_3$: the minimum number of locations in each sub-region is 2 and the maximum cost of the experiment is $50J$, with the costs of measuring in the five sub-regions being $\{40, 44, 50, 65,60\}$
  \item $C_4$: the minimum number of locations in each sub-region is 1, the maximum number of locations in each sub-region is $J/3$, and the maximum cost of the experiment is $50J$, with the costs of measuring in the five sub-regions being $\{40, 44, 50, 55, 60\}$, where the cost $55$ for the $4$-th sub-region is not the same as in case $C_3$.
\end{itemize}

The resulting exact designs after applying the constraints above, together with their MSE related values and their efficiencies with respect to the size-constrained designs are depicted in Table \ref{T:con}. Under the size-constrained designs we understand the designs with the only two restrictions of total number of locations $J$ and of at least one location in each sub-region. The optimal and highly efficient size-constrained designs were considered in the previous section and they are presented in Table~\ref{T:3k}. For simplicity, we will further call them unconstrained designs.

The efficiency $\mathrm{Eff}$ of the obtained constrained designs was computed by the following formula:
\begin{equation*}
\mathrm{Eff}=\Phi_{A,\alpha}(\xi^*)/\Phi_{A,\alpha}(\xi^*_c),
\end{equation*}
where $\xi^*$ and $\xi^*_c$ denote the unconstrained and the constrained designs, respectively. Obviously, the criterion values for unconstrained designs are smaller than or the same as for the constrained designs. Hence, the efficiency in always between $0$ and $1$. Here, it practically shows how much the two designs differ from each other.

The MSE related values denoted by $\mathrm{MSE}_{Tr}$ in the table coincide with the trace of MSE matrix \eqref{mse1}.

As we can see in Table~\ref{T:con}, in case $C_1$ the obtained constrained design differs from the unconstrained design (first block in Table~\ref{T:3k}, $K=80$) only for $J=20$. For $J=40$ and $J=100$, the unconstrained designs fulfill the additional restriction of at least two locations per sub-region. In case $C_2$, the obtained designs differ  for all values of $J$ from the unconstrained designs, for which the restriction of maximum $J/4$ locations per sub-region is not fulfilled. In case $C_3$, the constrained designs assign as expected more locations to the cheapest sub-regions $1$, and the number of locations in the most expensive sub-region $4$ is reduced. In case $C_4$, the obtained constrained design differs from the optimal unconstrained design only for $J=20$. For $J=40$ and $J=100$, the unconstrained designs satisfy the conditions of the maximal number of locations $J/3$ and the total cost of $50J$. For $J=20$, the obtained design assigns the minimal number of locations $1$ to the most expensive sub-region $5$. Since the largest number of locations per sub-region is bounded by $20/3=6.33$, in the cheapest sub-region $1$ the number of locations is $6$, and the design looks more balanced than in case $C_3$.

\subsection{Particular Case: Block-Diagonal Kinship Matrix with Compound Symmetry Blocks}\label{CS block-diag}


We consider an example for the particular case of block-diagonal kinship matrix with CS blocks that has been introduced in Section~\ref{part cases}. We fix the total number of locations by $J=40$, and we illustrate the behavior of optimal designs on several values of the covariance parameter and the number of genotypes: $r\in\{1/2, 1/3, 1/4\}$, $K\in\{30, 90, 300, 900\}$, and various combinations of the number of families $f$ and the number of genotypes per family $m$ (\citealt{piepho2006comparison}).  Although some of the combinations of $m$ and $f$ are rather irrelevant for practice (e.\,g. $m=150$ and $f=6$), they illustrate well the behavior of optimal designs and are, therefore, used in this example. The variance parameter $\sigma^2_\alpha$ is computed from the condition $\textrm{asv}(N)=1$ (\citealt{piepho2023adjusted}), 
where 
\begin{equation*}\label{asv}
    \textrm{asv}(N)=\frac{1}{K-1}\mathrm{tr}\left[\mathbf{N}\left(\mathbb{I}_K-\frac{1}{K}\mathds{1}_K\mathds{1}_K^\top\right)\right],
\end{equation*}
which is in accordance with the VanRaden method used in computing the kinship matrices for the example in Section~\ref{3 kinship matrices}.  For the block-diagonal structure of the kinship matrix $\mathbf{N}$ used in Theorem~\ref{CS2}, we obtain
\begin{equation*}
    \sigma^2_\alpha=\frac{K-1}{K-1-(m-1)r},
\end{equation*}
which is not smaller than $1$ for all values of $K$, $m$ and $r$, increases with increasing $m$ and $r$, and decreases with increasing $K$. In the trivial cases $m=1$ (one genotype per group) and $f=1$ (CS, i.\,e. only one block) we have, respectively, $\sigma^2_\alpha=1$ and $\sigma^2_\alpha=\frac{1}{1-r}$. With increasing number of genotypes $K$ the value of the variance $\sigma^2_\alpha$ approaches $1$.

We focus on the \textit{A}-criterion for model \eqref{mod}. For the weighted \textit{A}-criterion and model \eqref{mod2}, similar behavior of optimal designs can be observed. Since the dimension of the information matrix in the K-Bayesian criterion \eqref{kb1} is independent of the number of varieties and equal to the small value $2P$, we prefer to use the MISOCP procedure of the R OptimalDesign package for the computation of designs. For illustrative purposes we provide solutions for both optimal approximate and optimal exact designs. Note that the optimal exact designs received from MISOCP are not simply obtained from optimal approximate designs using rounding. These designs are truly optimal in the class of exact designs. 

Table~\ref{T:CS} illustrates the results. The MSE related values denoted by $\mathrm{MSE}_{Tr,a}$ and $\mathrm{MSE}_{Tr,e}$ coincide with the trace of MSE matrix \eqref{mse1} for approximate and exact designs, respectively. As we can see in the table, optimal designs depend on the values of $f$, $m$, $r$ and $K$. For all combinations, the optimal allocations of trials imply more locations in the sub-regions with larger variances. The latter behavior we also observed in the example considered in Section~\ref{3 kinship matrices}. The MSE values increase with increasing number of varieties per family $m$, the total number of varieties $K$, and the covariance parameter $r$, and they decrease with increasing number of families $f$. With respect to the number genotypes $K$ the behavior of the optimal designs considerably differs from that in Section~\ref{3 kinship matrices}, since even for very large values, like $K=900$, optimal designs are far from balanced. Hence, it may be very inefficient to use balanced designs in experiments with large number of genotypes.

\section{Discussion}

The main contribution compared to our previous work on the optimal allocation of trials to sub-regions (\citealt{pru10}) is that our new approach allows pedigree or kinship relations between genotypes to be taken into account at a scale in terms of the number of genotypes that is relevant in public and commercial breeding programs. In our experience, design generation with about 700 genotypes is feasible on a regular desktop (computations were performed on a 64-bit Windows 11 operating system running an AMD Ryzen 7 5800H CPU processor at 3.20 GHz with 16 GB RAM) with computing times in the order of 2 hours. One of the directions for the future research is the extension of the dimension reduction approach proposed in \cite{bodnar2026} for the present design problem. This approach can be combined with various computational methods for optimal designs, including the algorithm we used in the present paper, and, hence can allow for even larger numbers of genotypes, e.\,g, of the order of two-three thousand.

In the linear mixed model with uncorrelated genotype effects considered in \citealt{pru10}, the optimal allocations of trials for the prediction of genotype effects are the same as when the criterion is formulated in terms of their pairwise differences. By contrast, with the model allowing for correlated genotype effects as considered in the present work, the optimal design differs for these two cases. We believe that the criterion based on contrasts is the more relevant criterion in practice, because both breeders and variety testing authorities need to select the best genotypes, and selection is essentially based on pairwise comparisons. For this reason, we recommend the criterion based on differences for use in practice. The criterion proposed in this paper for the prediction of the genotype effects themselves constitutes a generalization of \cite{pru10}, and the general approach reduces to the one in \cite{pru10} if we replace the kinship with the identity matrix.

Optimization of a trialing network usually needs to be done under a number of constraints. The primary constraint in our approach is that the total number of trials, $J$, is given. Either $J$ is itself constrained to a specific value by the given budget, or it is a parameter to be optimized in another step, in which the value of $J$ achieving the desired precision for the overall network is determined. The search for this optimum can be implemented as a grid search that, e.\,g., gradually increases $J$ on a grid until the desired precision is achieved. One can also plot precision against $J$ and then determine the point at which the incremental improvement becomes negligible, or too small to justify the additional cost of running one more trial.

Our approach allows for additional linear constraints on the design to be imposed. Important options include minimum and maximum number of locations per sub-region in absolute values, maximum number of locations per sub-region related to the total number of locations, and the total cost constraint provided given sub-region specific costs of one location.


\section*{Acknowledgment} This work has been partially funded by the DFG-project PR-1615/3-1 and VEGA project no. 1/0480/26. We thank KWS SAAT AG (Dr. Carsten Knaak) for allowing us to use the marker data of Albrecht et al. (2011) on the first example. We further thank the INVITE project (https://www.h2020-invite.eu/), and the coordinators Claude Welcker and Karl Schmid in particular, for permitting us to use the project's two maize marker datasets.

\section*{Conflict of interest} There is no conflicts of interest to disclose.

\bibliographystyle{natbib}
\bibliography{prus10}

\newpage

\begin{table}[!ht] 
\caption{\small Optimal and highly efficient designs for standard and weighted \textit{A}-criteria for prediction of genotype effects and their pairwise linear contrasts in models~\eqref{mod} and \eqref{mod2} for three kinship matrices with $K=31$, $K=80$, and $K=698$ genotypes}\label{T:3k}
\vspace{0.2cm}
\hspace*{-1.5cm}
{\small
	\begin{tabular}{|c||ccccc|c||ccccc|c||ccccc|c|}
        \hline
          & \multicolumn{6}{c||}{$K=31$}  &  \multicolumn{6}{c||}{$K=80$}  &  \multicolumn{6}{c|}{$K=698$} \\
           \cline{2-19}
		$J$ & $J_1$ & $J_2$ & $J_3$ & $J_4$ & $J_5$ & $\mathrm{MSE}_{Tr}$ & $J_1$ & $J_2$ & $J_3$ & $J_4$ & $J_5$ & $\mathrm{MSE}_{Tr}$ & $J_1$ & $J_2$ & $J_3$ & $J_4$ & $J_5$ & $\mathrm{MSE}_{Tr}$ \\ 
        \hline
        \hline
        \multicolumn{19}{|c|}{ Standard \textit{A}-criterion for prediction of genotype effects in model~\eqref{mod}} \\
        \hline
		\hline
		10 & 4 & 1 & 1 & 3 & 1 & 6263& 4 & 1 & 1 & 3 & 1&17007 &2 & 2 & 2 & 2 & 2 & 1967370718\\ 
		20 & 7 & 2 & 3 & 7 & 1 & 4752& 7 & 2 & 3 & 7 & 1&12757 &4 & 4 & 4 & 4 & 4 & 1966922683\\ 
		40 & 13 & 6 & 7 & 12 & 2 &3745& 13 & 6 & 7 & 12 & 2&9935 &8 & 8 & 8 & 8 & 8 & 1966696245\\ 
		100 & 29 & 16 & 19 & 27 & 9 &2892& 28 & 17 & 20 & 27 & 8& 7513& 20 & 20 & 20 & 20 & 20 & 1966559528\\ 
		\hline
        \hline
        \multicolumn{19}{|c|}{ Standard \textit{A}-criterion for prediction of pairwise linear contrasts in model~\eqref{mod}} \\
        \hline
        \hline
    10 & 4 & 1 & 1 & 3 & 1 &18705& 4 & 1 & 1 & 3 & 1&1360577& 2 & 2 & 2 & 2 & 2 & 365289004\\ 
	20 & 7 & 2 & 3 & 7 & 1 &141076& 7 & 2 & 3 & 7 & 1&1020581& 4 & 4 & 4 & 4 & 4 & 203377337\\ 
	40 & 13 & 6 & 7 & 12 & 2&109846& 13 & 6 & 7 & 12 & 2 &794832& 9 & 8 & 7 & 8 & 8 &122289244\\ 
	100 & 29 & 16 & 19 & 27 & 9&83434& 28 & 17 & 20 & 27 & 8&601071 &22 & 18 & 18 & 22 & 20 & 72854945 \\ 
    \hline
        \hline
        \multicolumn{19}{|c|}{ Weighted \textit{A}-criterion for prediction of genotype effects in model~\eqref{mod}} \\
        \hline
        \hline
	10 & 3 & 1 & 1 & 4 & 1&6263 &3 & 1 & 1 & 4 & 1 &17007&2 & 2 & 2 & 2 & 2 & 1967370718\\ 
	20 & 8 & 1 & 2 & 8 & 1&4799 & 8 & 1 & 2 & 8 & 1& 12879&4& 3 & 3 & 5 & 5 & 1966947073\\ 
	40 & 13 & 4 & 5 & 14 & 4& 3805& 13 & 4 & 5 & 14 & 4& 10096& 8 & 6 & 7 & 9 &  10 & 1966703841\\ 
	100 & 30 & 11 & 15 & 31 & 13&2922 & 29 & 12 & 15 & 31 & 13&7597 & 21 &15 &  16& 23 &25 & 1966563308\\ 
        \hline
        \hline
        \multicolumn{19}{|c|}{ Weighted \textit{A}-criterion for prediction of pairwise linear contrasts in model~\eqref{mod}} \\
        \hline
        \hline
    10 & 3 & 1 & 1 & 4 & 1 &187905 &3 & 1 & 1 & 4 & 1&1360577 &2 & 2 & 2 & 2 & 2& 365289004\\ 
	20 & 8 & 1 & 2 & 8 & 1 &142539 &8 & 1 & 2 & 8 & 1&1030296 &4 & 4 & 3 & 5 & 4 & 207720975\\ 
	40 & 13 & 4 & 5 & 14 & 4&1111703 & 13 & 4 & 5 & 14 & 4& 807687& 8 & 6 & 7 & 9 & 10 & 124499337\\ 
	100 & 30 & 11 & 15 & 31 & 13& 84343& 29 & 12 & 15 & 31 & 13&606427 & 18 & 18& 20 & 22& 22 & 72860376\\ 
    \hline
    \hline
    \multicolumn{19}{|c|}{ Standard \textit{A}-criterion for prediction of genotype effects in model~\eqref{mod2}} \\
    \hline
    \hline
    		10 & 4 & 1 & 1 & 3 & 1 & 5160& 4 & 1 & 1 & 3 & 1&13890 &2 & 2 & 2 & 2 & 2 & 1967018553\\ 
		20 & 7 & 3 & 3 & 6 & 1 & 3995& 7 & 3 & 3 & 6 & 1&10639 &4 & 4 & 4 & 4 & 4 & 1966744572\\ 
		40 & 12 & 6 & 7 & 12 & 3 &3225& 12 & 6 & 8 & 12 & 2&8456 &8 & 8 & 8 & 8 & 8 & 1966606590\\ 
		100 & 28 & 17 & 19 & 26 & 10 &2596& 27 & 17 & 20 & 26 & 10& 6671& 20 & 20 & 20 & 20 & 20 & 1966523569\\ 
		\hline
        \hline
        \multicolumn{19}{|c|}{ Standard \textit{A}-criterion for prediction of pairwise linear contrasts in model~\eqref{mod2}} \\
        \hline
        \hline
    	10 & 4 & 1 & 1 & 3 & 1 &153729& 4 & 1 & 1 & 3 & 1&1111193& 2 & 2 & 2 & 2 & 2 & 237972878\\ 
	20 & 7 & 3 & 3 & 6 & 1 &117605& 7 & 3 & 3 & 6 & 1&851113& 4 & 4 & 4 & 4 & 4& 139175319\\ 
	40 & 12 & 6 & 7 & 12 & 3&93747& 12 & 6 & 8 & 12 & 2 &676512& 8 & 8 & 8 & 8 & 8 & 89528399\\ 
	100 & 28 & 17 & 19 & 26 & 10&74257& 27 & 17 & 20 & 26 & 10&533724 &20 & 20 & 20 & 20 & 20 & 59658022\\ 
    \hline
        \hline
        \multicolumn{19}{|c|}{ Weighted \textit{A}-criterion for prediction of genotype effects in model~\eqref{mod2}} \\
        \hline
        \hline
	10 & 3 & 1 & 1 & 4 & 1&5180 &3 & 1 & 1 & 4 & 1 &13942&2 & 2 & 2 & 2 & 2 & 1967018553\\ 
	20 & 7 & 2 & 2 & 8 & 1&4044 & 7 & 2 & 2 & 8 & 1& 10772&4& 3 & 3 & 5 & 5 & 1966759381\\ 
	40 & 13 & 4 & 5 & 14 & 4& 3270& 13 & 4 & 6 & 13 & 4& 8546& 8 & 6 & 7 & 9 &  10 & 1966611243\\ 
	100 & 29 & 12 & 15 & 31 & 13&2614 & 29 & 12 & 15 & 31 & 13&6730 & 18 &17 &  18& 20 &27 & 1966525133\\ 
            \hline
        \hline
        \multicolumn{19}{|c|}{ Weighted \textit{A}-criterion for prediction of pairwise linear contrasts in model~\eqref{mod2}} \\
        \hline
        \hline
	10 & 3 & 1 & 1 & 4 & 1 &154345 &3 & 1 & 1 & 4 & 1&1115357 &2 & 2 & 2 & 2 & 2& 237972878\\ 
	20 & 7 & 2 & 2 & 8 & 1 &119124 &7 & 2 & 2 & 8 & 1&861745 &4 & 3 & 3 & 5 & 5& 144497239\\ 
	40 & 13 & 4 & 5 & 14 & 4&95137 & 13 & 4 & 6 & 13 & 4& 683709& 8 & 6 & 7 & 9 & 10& 91180712 \\ 
	100 & 29 & 12 & 15 & 31 & 13& 74790& 29 & 12 & 15 & 31 & 13&538479 & 21 & 15& 16 & 24 & 24& 60466310\\ 
    \hline
	\end{tabular}
    }
\end{table}

\newpage
\begin{table}[!ht]
\caption{Optimal and highly efficient designs for standard \textit{A}-criterion in model (1) for 80 genotypes under additional constraint sets $C_1-C_4$}\label{T:con}
\vspace{0.2cm}
\centering
\begin{tabular}{|c||ccccc|c|c||ccccc|c|c|}
\hline
	$J$ & $J_1$ & $J_2$ & $J_3$ & $J_4$ & $J_5$ & $\mathrm{Eff}$ & $\mathrm{MSE}_{Tr}$ & $J_1$ & $J_2$ & $J_3$ & $J_4$ & $J_5$ & $\mathrm{Eff}$ & $\mathrm{MSE}_{Tr}$ \\
    \hline \hline
    & \multicolumn{7}{c||}{$C_1$}  &  \multicolumn{7}{c|}{$C_2$} \\
\hline \hline
20 & 7  & 2  & 3  & 6  & 2  & 0.99 &12866& 5  & 3  & 5  & 5  & 2  & 0.94 & 13278\\
40 & 13 & 6  & 7  & 12 & 2  & 1.00 &9935& 10 & 8  & 9  & 10 & 3  & 0.97 & 10104\\
100 & 28 & 17 & 20 & 27 & 8  & 1.00 &7513& 25 & 19 & 22 & 25 & 9  & 0.99 & 7538\\
\hline \hline
& \multicolumn{7}{c||}{$C_3$}  &  \multicolumn{7}{c|}{$C_4$} \\
\hline \hline
20 &9  & 2  & 2  & 5  & 2  & 0.97 &13013& 6  & 3  & 4  & 6  & 1  & 0.97&12904\\
40&15 & 6  & 6  & 11 & 2  & 0.99 &9972& 13 & 6  & 7  & 12 & 2  & 1.00&9935\\
100&33 & 17 & 19 & 24 & 7  & 0.99 &7531& 29 & 16 & 19 & 28 & 8  & 1.00&7513\\
\hline
\end{tabular}
\end{table}

\newpage
\begin{table}[!ht] 
\caption{Optimal approximate and exact designs for standard \textit{A}-criterion in model (1) for block-diagonal kinship matrix with CS blocks}\label{T:CS}
\vspace{0.2cm}
\centering
\begin{tabular}{|c|c|c|ccccc|c|ccccc|c|}
\hline
 &  & & \multicolumn{5}{c|}{Approximate design} &
& \multicolumn{5}{c|}{Exact design} &
\\
 $r$ & $f$ & $m$ &  $w_1$ & $w_2$ & $w_3$ & $w_4$ & $w_5$ & $\mathrm{MSE}_{Tr,a}$ & $J_1$ & $J_2$ & $J_3$ & $J_4$ & $J_5$ & $\mathrm{MSE}_{Tr,e}$\\
\hline
\hline
\multicolumn{15}{|c|}{$K=30$} \\
\hline
\hline
1/2 & 6 & 5 & 0.33 & 0.14 & 0.18 & 0.31 & 0.04 & 8751 & 13 & 6 & 7 & 13 & 1 & 8752 \\
\hline
& 15 & 2 & 0.33 & 0.14 & 0.19 & 0.31 & 0.03 & 6265 & 13 & 6 & 8 & 12 & 1 & 6266 \\
\hline
\hline
1/3 & 6 & 5 & 0.33 & 0.14 & 0.19 & 0.32 & 0.02 & 7708 & 13 & 6 & 7 & 13 & 1 & 7709 \\
\hline
& 15 & 2 & 0.33 & 0.14 & 0.19 & 0.31 & 0.03 & 6070 & 13 & 6 & 8 & 12 & 1 & 6071 \\
\hline
\hline
1/4 & 6 & 5 & 0.33 & 0.14 & 0.19 & 0.32 & 0.02 & 7181 & 13 & 6 & 7 & 13 & 1 & 7183 \\
\hline
& 15 & 2 & 0.33 & 0.14 & 0.19 & 0.31 & 0.03 & 5958 & 13 & 6 & 8 & 12 & 1 & 5960 \\
\hline
\hline
\multicolumn{15}{|c|}{$K=90$} \\
\hline
\hline
1/2 & 6 & 15 & 0.34 & 0.12 & 0.16 & 0.32 & 0.06 & 24359 & 14 & 5 & 6 & 13 & 2 & 24361 \\
\hline
& 18 & 5 & 0.33 & 0.14 & 0.18 & 0.31 & 0.04 & 16040 & 13 & 6 & 7 & 13 & 1 & 16044  \\
\hline
\hline
1/3 & 6 & 15 & 0.33 & 0.13 & 0.18 & 0.32 & 0.04 & 20897 & 13 & 5 & 7 & 13 & 2 & 20903  \\
\hline
& 18 & 5 & 0.33 & 0.14 & 0.19 & 0.31 & 0.03 & 15513 & 13 & 6 & 7 & 13 & 1 & 15516  \\
\hline
\hline
1/4 & 6 & 15 & 0.33 & 0.14 & 0.19 & 0.32 & 0.02 & 19155 & 13 & 6 & 7 & 13 & 1 & 19159 \\
\hline
& 18 & 5 & 0.33 & 0.14 & 0.19 & 0.32 & 0.02 & 15173 & 13 & 6 & 7 & 13 & 1 & 15178  \\
\hline
\hline
\multicolumn{15}{|c|}{$K=300$} \\
\hline
\hline
1/2 & 6 & 50 & 0.36 & 0.10 & 0.15 & 0.34 & 0.05 & 78247 & 14 & 4 & 6 & 14 & 2 & 78253\\
\hline
& 15 & 20 & 0.35 & 0.12 & 0.15 & 0.32 & 0.06 & 52068 & 14 & 5 & 6 & 13 & 2 & 52082 \\
\hline
& 60 & 5 & 0.33 & 0.14 & 0.18 & 0.31 & 0.04 & 42327 & 13 & 6 & 7 & 13 & 1 & 42343 \\
\hline
\hline
1/3 & 6 & 50 & 0.34 & 0.12 & 0.17 & 0.33 & 0.04 & 66473 & 13 & 5 & 7 & 13 & 2 & 66490 \\
\hline
& 15 & 20 & 0.34 & 0.13 & 0.17 & 0.32 & 0.04 & 49776 & 13 & 5 & 7 & 13 & 2 & 49787 \\
\hline
& 60 & 5 & 0.33 & 0.14 & 0.19 & 0.31 & 0.03 & 43215 & 13 & 6 & 7 & 13 & 1 & 43229 \\
\hline
\hline
1/4 & 6 & 50 & 0.34 & 0.12 & 0.18 & 0.32 & 0.04 & 60588 & 14 & 5 & 7 & 13 & 1 & 60607 \\
\hline
& 15 & 20 & 0.33 & 0.14 & 0.18 & 0.32 & 0.03 & 48349 & 14 & 5 & 7 & 13 & 1 & 48366 \\
\hline
& 60 & 5 & 0.33 & 0.14 & 0.19 & 0.31 & 0.03 & 43344 & 13 & 6 & 8 & 12 & 1 & 43408 \\
\hline
\hline
\multicolumn{15}{|c|}{$K=900$} \\
\hline
\hline
1/2 & 6 & 150 & 0.38 & 0.08 & 0.14 & 0.36 & 0.04 & 231418 & 14 & 4 & 6 & 15 & 1 & 231627\\
\hline
& 15 & 60 & 0.37 & 0.10 & 0.14 & 0.34 & 0.05 & 151087 & 15 & 4 & 6 & 13 & 2 & 151125\\
\hline
& 60 & 15 & 0.35 & 0.12 & 0.15 & 0.32 & 0.06 & 118676 & 14 & 5 & 6 & 13 & 2 & 118707\\
\hline
\hline
1/3 & 6 & 150 & 0.35 & 0.09 & 0.17 & 0.35 & 0.04 & 195903 & 14 & 4 & 7 & 14 & 1 & 196062\\
\hline
& 15 & 60 & 0.35 & 0.11 & 0.16 & 0.33 & 0.05 & 144757 & 14 & 4 & 7 & 13 & 2 & 144805\\
\hline
& 60 & 15 & 0.34 & 0.13 & 0.17 & 0.32 & 0.04 & 123656 & 13 & 5 & 7 & 13 & 2 & 123715\\
\hline
\hline
1/4 & 6 & 150 & 0.34 & 0.12 & 0.18 & 0.33 & 0.03 & 178305 & 14 & 5 & 7 & 13 & 1 & 178343\\
\hline
& 15 & 60 & 0.35 & 0.12 & 0.15 & 0.32 & 0.06 & 140810 & 14 & 5 & 7 & 12 & 2 & 140899\\
\hline
& 60 & 15 & 0.33 & 0.14 & 0.18 & 0.32 & 0.03 & 125205 & 13 & 6 & 7 & 13 & 1 & 125245\\
\hline
\end{tabular}
\end{table}

\newpage

\appendix

{\Large \textbf{Web Appendices}}

\section{Proof of Lemma~\ref{lem1}}\label{pl1}

    From the proof of Lemma~1 in \cite{pru10} it follows that
    \begin{equation*}
\mathrm{Cov}(\hat{\mbox{\boldmath{$\alpha$}}}-\mbox{\boldmath{$\alpha$}})=\left[\left(\mathbb{I}_K-\frac{1}{K}\mathds{1}_K\mathds{1}_K^\top\right)\otimes\left(H\,\mathbf{F}^\top\mathbf{W}_2^{-1}\mathbf{F}\right)+\mathbf{U}^{-1}\right]^{-1},
\end{equation*}
where
\begin{equation*}
\mathbf{W}_2=\sigma^2_\omega\mathds{1}_J\mathds{1}_J^\top+\sigma^2_\tau\mathbf{F}\mathbf{F}^\top+c\,H\,\mathbb{I}_J,
\end{equation*}
and that
\begin{equation*}
 \mathbf{F}^\top\left(\frac{\sigma^2_\omega}{H}\mathds{1}_J\mathds{1}_J^\top+\frac{\sigma^2_\tau}{H}\mathbf{F}\mathbf{F}^\top+c\,\mathbb{I}_J\right)^{-1}\mathbf{F} =
\frac{H}{\sigma^2_\tau}\left\{\mathbb{I}_P+\frac{\sigma^2_\omega}{\sigma^2_\tau}\mathds{1}_P\mathds{1}_P^\top+\frac{c\,H}{\sigma^2_\tau}\left(\mathbf{F}^\top\mathbf{F}\right)^{-1}\right\}^{-1}.
\end{equation*}
Then from 
\begin{equation*}
 H\,\mathbf{F}^\top\mathbf{W}_2^{-1}\mathbf{F}=\mathbf{F}^\top\left(\frac{\sigma^2_\omega}{H}\mathds{1}_J\mathds{1}_J^\top+\frac{\sigma^2_\tau}{H}\mathbf{F}\mathbf{F}^\top+c\,\mathbb{I}_J\right)^{-1}\mathbf{F}
\end{equation*}
we obtain
\begin{equation*}
 H\,\mathbf{F}^\top\mathbf{W}_2^{-1}\mathbf{F}=\frac{1}{c}\left\{\frac{\sigma^2_\tau}{c\,H}\mathbb{I}_P+\frac{\sigma^2_\omega}{c\,H}\mathds{1}_P\mathds{1}_P^\top+\left(\mathbf{F}^\top\mathbf{F}\right)^{-1}\right\}^{-1},
\end{equation*}
which implies the result of the lemma.

\section{Proof of Theorem~\ref{t3}}\label{pt1}

From the definition of the criterion and formula \eqref{mse1} it follows that
\begin{equation*}
\mathrm{tr}\left(\mathrm{MSE}_{\alpha}(\xi)\left(\mathbb{I}_K\otimes\mathcal{L}\right)\right)=\mathrm{tr}\left(\left\{\mathbf{T}\otimes\left[\mathbf{M}(\xi)^{-1}+\tilde{\mathbf{R}}\right]^{-1}+\tilde{\mathbf{U}}^{-1}\right\}^{-1}\left(\mathbb{I}_K\otimes\mathcal{L}\right)\right).
\end{equation*}
Using the Woodbury formula (see e.\,g. \citealt{Harville97}, p. 424), we obtain
\begin{equation*}
\left\{\mathbf{T}\otimes\left[\mathbf{M}(\xi)^{-1}+\tilde{\mathbf{R}}\right]^{-1}+\tilde{\mathbf{U}}^{-1}\right\}^{-1}=\tilde{\mathbf{U}}-\tilde{\mathbf{U}}\left(\mathbf{T}\otimes\mathbb{I}_P\right)\left[
\mathbb{I}_K\otimes\mathbf{M}(\xi)^{-1}+\mathbf{W}\right]^{-1}\left(\mathbf{T}\otimes\mathbb{I}_P\right)\tilde{\mathbf{U}},
\end{equation*}
where $\mathbf{W}=\mathbb{I}_K\otimes\tilde{\mathbf{R}}+\left(\mathbf{T}\otimes\mathbb{I}_P\right)\tilde{\mathbf{U}}\left(\mathbf{T}\otimes\mathbb{I}_P\right)$. From the standard inversion formula $(A_1+A_2)^{-1}=A_1^{-1}-A_1^{-1}(A_1^{-1}+A_2^{-1})^{-1}A_1^{-1}$ for two non-singular matrices $A_1$ and $A_2$, we receive
\begin{equation*}
\left[
\mathbb{I}_K\otimes\mathbf{M}(\xi)^{-1}+\mathbf{W}\right]^{-1}=\mathbf{W}^{-1}-\mathbf{W}^{-1}\left[
\mathbb{I}_K\otimes\mathbf{M}(\xi)+\mathbf{W}^{-1}\right]^{-1}\mathbf{W}^{-1},
\end{equation*}
which implies for $\mathbf{B}=\mathbf{W}^{-1}$
\begin{equation}\label{mse4}
\left\{\mathbf{T}\otimes\left[\mathbf{M}(\xi)^{-1}+\tilde{\mathbf{R}}\right]^{-1}+\tilde{\mathbf{U}}^{-1}\right\}^{-1}=\mathbf{C}_1+\tilde{\mathbf{U}}\left(\mathbf{T}\otimes\mathbb{I}_P\right)\mathbf{B}\left[
\mathbb{I}_K\otimes\mathbf{M}(\xi)+\mathbf{B}\right]^{-1}\mathbf{B}\left(\mathbf{T}\otimes\mathbb{I}_P\right)\tilde{\mathbf{U}},
\end{equation}
where $\mathbf{C}_1=\tilde{\mathbf{U}}-\tilde{\mathbf{U}}\left(\mathbf{T}\otimes\mathbb{I}_P\right)\mathbf{B}\left(\mathbf{T}\otimes\mathbb{I}_P\right)\tilde{\mathbf{U}}$.
Then 
\begin{eqnarray*}
\mathrm{tr}\left(\mathrm{MSE}_{\alpha}(\xi)\left(\mathbb{I}_K\otimes\mathcal{L}\right)\right)&=&c_1+
\mathrm{tr}\left\{\tilde{\mathbf{U}}\left(\mathbf{T}\otimes\mathbb{I}_P\right)\mathbf{B}\left[
\mathbb{I}_K\otimes\mathbf{M}(\xi)+\mathbf{B}\right]^{-1}\mathbf{B}\left(\mathbf{T}\otimes\mathbb{I}_P\right)\tilde{\mathbf{U}}\left(\mathbb{I}_K\otimes\mathcal{L}\right)\right\} \\
&=&c_1+
\mathrm{tr}\left\{\left[
\mathbb{I}_K\otimes\mathbf{M}(\xi)+\mathbf{B}\right]^{-1}\mathbf{B}\left(\mathbf{T}\otimes\mathbb{I}_P\right)\tilde{\mathbf{U}}\left(\mathbb{I}_K\otimes\mathcal{L}\right)\tilde{\mathbf{U}}\left(\mathbf{T}\otimes\mathbb{I}_P\right)\mathbf{B}\right\},
\end{eqnarray*}
where $c_1=\mathrm{tr}[\mathbf{C}_1\left(\mathbb{I}_K\otimes\mathcal{L}\right)]$ has no influence on design.

\section{Proof of Corollary~\ref{t4}}\label{pc3}

For the contrasts matrix $\mathcal{C}$ it holds that
\begin{equation*}
\mathcal{C}^\top \mathcal{C}=\sum_{s=1}^{K-1}\mathcal{C}_s^\top \mathcal{C}_s=K\mathbb{I}_K-\mathds{1}_K\mathds{1}_K^\top=K\,\mathbf{T},
\end{equation*}
since  $\mathbf{T}=\mathbb{I}_K-\frac{1}{K}\mathds{1}_K\mathds{1}_K^\top$.
Then, 
\begin{eqnarray*}
    \mathrm{tr}\left[\left(\mathcal{C}\otimes\mathbb{I}_P\right)\mathrm{MSE}_{\alpha}(\xi)\left(\mathcal{C}^\top\otimes\mathbb{I}_P\right)\left(\mathbb{I}_n\otimes\mathcal{L}\right)\right]
    &=&\mathrm{tr}\left\{\mathrm{MSE}_{\alpha}(\xi)\left[\left(\mathcal{C}^\top\mathcal{C}\right)\otimes\mathcal{L}\right]\right\}\\
    &=&K\,\mathrm{tr}\left\{\mathrm{MSE}_{\alpha}(\xi)\left(\mathbf{T}\otimes\mathcal{L}\right)\right\}.
\end{eqnarray*} 

From formula \eqref{mse4} in the proof of Theorem~\ref{t3} (see Web Appendix~\ref{pt1}), we obtain
\begin{eqnarray*}
    \mathrm{tr}\left\{\mathrm{MSE}_{\alpha}(\xi)\left(\mathbf{T}\otimes\mathcal{L}\right)\right\}&=&c_2+
\mathrm{tr}\left\{\left[
\mathbb{I}_K\otimes\mathbf{M}(\xi)+\mathbf{B}\right]^{-1}\mathbf{B}\left(\mathbf{T}\otimes\mathbb{I}_P\right)\tilde{\mathbf{U}}\left(\mathbf{T}\otimes\mathcal{L}\right)\tilde{\mathbf{U}}\left(\mathbf{T}\otimes\mathbb{I}_P\right)\mathbf{B}\right\},
\end{eqnarray*} 
where the additive term $c_2=\mathrm{tr}[\mathbf{C}_1\left(\mathbf{T}\otimes\mathcal{L}\right)]$ is independent of $\xi$ and, hence, has no influence on design.

\section{Proof of Theorem~\ref{CS1}}\label{pt2}

The result will be verified for the weighted \textit{A}-criterion only, since the standard \textit{A}-criterion is a particular case of it.

For $\mathbf{N}=a_1\mathbb{I}_K+a\mathds{1}_K\mathds{1}_K^\top$ it holds that $\mathbf{N}^{-1}=\frac{1}{a_1}\left(\mathbb{I}_K-u\frac{1}{K}\mathds{1}_K\mathds{1}_K^\top\right)$, where $u=\frac{aK}{aK+a_1}$. Then the MSE matrix \eqref{mse1} can be written as
\begin{eqnarray}
        \mathrm{MSE}_{\alpha}(\xi)&=&\left\{(\mathbb{I}_K-\frac{1}{K}\mathds{1}_K\mathds{1}_K^\top)\otimes[\mathbf{M}(\xi)^{-1}+\tilde{\mathbf{R}}]^{-1}+(\mathbb{I}_K-u\frac{1}{K}\mathds{1}_K\mathds{1}_K^\top)\otimes\left(\frac{1}{a_1}\tilde{\mathbf{V}}^{-1}\right)\right\}^{-1}\nonumber \\
        &=&\left\{(\mathbb{I}_K-\frac{1}{K}\mathds{1}_K\mathds{1}_K^\top)\otimes\left\{[\mathbf{M}(\xi)^{-1}+\tilde{\mathbf{R}}]^{-1}+\frac{1}{a_1}\tilde{\mathbf{V}}^{-1}\right\}+\frac{1}{K}\left(\mathds{1}_K\mathds{1}_K^\top\right)\otimes\left(\frac{1-u}{a_1}\tilde{\mathbf{V}}^{-1}\right)\right\}^{-1}\nonumber \\
             &=&\left\{(\mathbb{I}_K-\frac{1}{K}\mathds{1}_K\mathds{1}_K^\top)\otimes\left\{[\mathbf{M}(\xi)^{-1}+\tilde{\mathbf{R}}]^{-1}+\frac{1}{a_1}\tilde{\mathbf{V}}^{-1}\right\}^{-1}+\mathbf{C}\right\}\label{msecs},
\end{eqnarray}
where $\mathbf{C}=\frac{1}{K}\mathds{1}_K\mathds{1}_K^\top\otimes\left(\frac{a_1}{1-u}\tilde{\mathbf{V}}\right)$.
Using the standard inversion formula $(A_1+A_2)^{-1}=A_1^{-1}-A_1^{-1}(A_1^{-1}+A_2^{-1})^{-1}A_1^{-1}$ for two non-singular matrices $A_1$ and $A_2$, we obtain
\begin{eqnarray}
    \left\{[\mathbf{M}(\xi)^{-1}+\tilde{\mathbf{R}}]^{-1}+\frac{1}{a_1}\tilde{\mathbf{V}}^{-1}\right\}^{-1}&=&a_1\tilde{\mathbf{V}}-a_1^2\tilde{\mathbf{V}}\left[\mathbf{M}^{-1}(\xi)+\mathbf{S}\right]^{-1}\tilde{\mathbf{V}}\nonumber \\
    &=&a_1\tilde{\mathbf{V}}-a_1^2\tilde{\mathbf{V}}\left\{\mathbf{S}^{-1}-\mathbf{S}^{-1}\left[\mathbf{M}(\xi)+\mathbf{S}^{-1}\right]^{-1}\mathbf{S}^{-1}\right\}\tilde{\mathbf{V}}\nonumber \\
    &=& a_1(\tilde{\mathbf{V}}-a_1\tilde{\mathbf{V}}\mathbf{S}^{-1}\tilde{\mathbf{V}})+a_1^2\tilde{\mathbf{V}}\mathbf{S}^{-1}\left[\mathbf{M}(\xi)+\mathbf{S}^{-1}\right]^{-1}\mathbf{S}^{-1}\tilde{\mathbf{V}}\label{eq1}
\end{eqnarray}
Then ti follows from \eqref{msecs} and \eqref{eq1} that 
\begin{eqnarray}
\mathrm{tr}\left[\mathrm{MSE}_{\alpha}(\xi)\left(\mathbb{I}_K\otimes\mathcal{L}\right)\right]&=&a_1^2(K-1)\mathrm{tr}\left\{\tilde{\mathbf{V}}\mathbf{S}^{-1}\left[\mathbf{M}(\xi)+\mathbf{S}^{-1}\right]^{-1}\mathbf{S}^{-1}\tilde{\mathbf{V}}\mathcal{L}\right\}+const\nonumber \\
&=&a_1^2(K-1)\mathrm{tr}\left\{\left[\mathbf{M}(\xi)+\mathbf{S}^{-1}\right]^{-1}\mathbf{S}^{-1}\tilde{\mathbf{V}}\mathcal{L}\tilde{\mathbf{V}}\mathbf{S}^{-1}\right\}+const,\label{msetrcs1}
\end{eqnarray}
where $const=a_1(K-1)\mathrm{tr}\left[\left(\tilde{\mathbf{V}}-a_1\tilde{\mathbf{V}}\mathbf{S}^{-1}\tilde{\mathbf{V}}\right)\mathcal{L}\right]+\mathrm{tr}\left(\mathbf{C}\mathcal{L}\right)$.

Matrix $\mathbf{T}=\mathbb{I}_K-\frac{1}{K}\mathds{1}_K\mathds{1}_K^\top$ is idempotent and it holds that $\left(\mathds{1}_K\mathds{1}_K^\top\right)\mathbf{T}=\mathbf{0}$. Hence, 
\begin{eqnarray}
\mathrm{tr}\left[\mathrm{MSE}_{\theta}(\xi)\left(\mathbb{I}_K\otimes\mathcal{L}\right)\right]
&=&a_1^2(K-1)\mathrm{tr}\left\{\left[\mathbf{M}(\xi)+\mathbf{S}^{-1}\right]^{-1}\mathbf{S}^{-1}\tilde{\mathbf{V}}\mathcal{L}\tilde{\mathbf{V}}\mathbf{S}^{-1}\right\}+const_1,\label{msetrcs2}
\end{eqnarray}
where $const_1=a_1(K-1)\mathrm{tr}\left[\left(\tilde{\mathbf{V}}-a_1\tilde{\mathbf{V}}\mathbf{S}^{-1}\tilde{\mathbf{V}}\right)\mathcal{L}\right]$. The constant terms $const$ and $const_1$, and the constant multiplicator $a_1^2(K-1)$ have no influence on designs. The weighted \textit{A}-criteria \eqref{wacrappr} and \eqref{wacrapprdef} differ from \eqref{msetrcs1} and \eqref{msetrcs2}, respectively, only by these constants.

\section{Proof of Theorem~\ref{CS2}}\label{pt3}

This result we prove for the weighted \textit{A}-criterion only, as the standard \textit{A}-criterion is the particular case of it with $\mathcal{L}=\mathbb{I}_P$.

The inverse of the kinship matrix $\mathbf{N}$ is given by $\mathbf{N}^{-1}=\frac{1}{b_1}\mathbb{I}_f\otimes\left(\mathbb{I}_m-v\frac{1}{m}\mathds{1}_m\mathds{1}_m^\top\right)$. Then, the the MSE matrix \eqref{mse1} can be written as

\begin{eqnarray*}
        \hspace{-1cm}\mathrm{MSE}_{\alpha}(\xi)&=&\left\{(\mathbb{I}_f\otimes\mathbb{I}_m-\frac{1}{f}\mathds{1}_f\mathds{1}_f^\top\otimes\frac{1}{m}\mathds{1}_m\mathds{1}_m^\top)\otimes[\mathbf{M}(\xi)^{-1}+\tilde{\mathbf{R}}]^{-1}+\mathbb{I}_f\otimes(\mathbb{I}_m-v\frac{1}{m}\mathds{1}_m\mathds{1}_m^\top)\otimes\left(\frac{1}{b_1}\tilde{\mathbf{V}}^{-1}\right)\right\}^{-1} \\
        &=&\left\{\mathbb{I}_f\otimes\left[\mathbb{I}_m\otimes\left(\mathbf{M}(\xi)^{-1}+\tilde{\mathbf{R}}\right)^{-1}+\left(\mathbb{I}_m-v\frac{1}{m}\mathds{1}_m\mathds{1}_m^\top\right)\otimes\left(\frac{1}{b_1}\tilde{\mathbf{V}}^{-1}\right)\right]\right.\\
        &&\hspace{7cm}-\left.\frac{1}{f}\mathds{1}_f\mathds{1}_f^\top\otimes\frac{1}{m}\mathds{1}_m\mathds{1}_m^\top\otimes\left(\mathbf{M}(\xi)^{-1}+\tilde{\mathbf{R}}\right)^{-1}\right\}^{-1} \\
             &=&\left(\mathbb{I}_f\otimes\left\{\mathbb{I}_m\otimes\left[\left(\mathbf{M}(\xi)^{-1}+\tilde{\mathbf{R}}\right)^{-1}+\frac{1}{b_1}\tilde{\mathbf{V}}^{-1}\right]-\frac{1}{m}\mathds{1}_m\mathds{1}_m^\top\otimes\left(\frac{1}{b_1}\tilde{\mathbf{V}}^{-1}\right)\right\}\right.\\
             &&\hspace{7cm}-\left.\frac{1}{f}\mathds{1}_f\mathds{1}_f^\top\otimes\frac{1}{m}\mathds{1}_m\mathds{1}_m^\top\otimes\left(\mathbf{M}(\xi)^{-1}+\tilde{\mathbf{R}}\right)^{-1}\right)^{-1}\\
             &=&\left(\mathbb{I}_f-\frac{1}{f}\mathds{1}_f\mathds{1}_f^\top\right)\otimes\left\{\mathbb{I}_m\otimes\left[\left(\mathbf{M}(\xi)^{-1}+\tilde{\mathbf{R}}\right)^{-1}+\frac{1}{b_1}\tilde{\mathbf{V}}^{-1}\right]-\frac{1}{m}\mathds{1}_m\mathds{1}_m^\top\otimes\left(\frac{v}{b_1}\tilde{\mathbf{V}}^{-1}\right)\right\}^{-1}\\
             &&+\frac{1}{f}\mathds{1}_f\mathds{1}_f^\top\otimes\left\{\mathbb{I}_m\otimes\left[\left(\mathbf{M}(\xi)^{-1}+\tilde{\mathbf{R}}\right)^{-1}+\frac{1}{b_1}\tilde{\mathbf{V}}^{-1}\right]-\frac{1}{m}\mathds{1}_m\mathds{1}_m^\top\otimes\left[\left(\mathbf{M}(\xi)^{-1}+\tilde{\mathbf{R}}\right)^{-1}+\frac{v}{b_1}\tilde{\mathbf{V}}^{-1}\right]\right\}^{-1}.
\end{eqnarray*}
Further we obtain
\begin{eqnarray*}
&&\hspace{-1cm}\left\{\mathbb{I}_m\otimes\left[\left(\mathbf{M}(\xi)^{-1}+\tilde{\mathbf{R}}\right)^{-1}+\frac{1}{b_1}\tilde{\mathbf{V}}^{-1}\right]-\frac{1}{m}\mathds{1}_m\mathds{1}_m^\top\otimes\left(\frac{v}{b_1}\tilde{\mathbf{V}}^{-1}\right)\right\}^{-1}\\
\quad &=& \left(\mathbb{I}_m-\frac{1}{m}\mathds{1}_m\mathds{1}_m^\top\right)\otimes\left[\left(\mathbf{M}(\xi)^{-1}+\tilde{\mathbf{R}}\right)^{-1}+\mathbf{V}_1^{-1}\right]^{-1}+\frac{1}{m}\mathds{1}_m\mathds{1}_m^\top\otimes\left[\left(\mathbf{M}(\xi)^{-1}+\tilde{\mathbf{R}}\right)^{-1}+\mathbf{V}_2^{-1}\right]^{-1}
\end{eqnarray*}
and 
\begin{eqnarray*}
&&\hspace{-1cm}\left\{\mathbb{I}_m\otimes\left[\left(\mathbf{M}(\xi)^{-1}+\tilde{\mathbf{R}}\right)^{-1}+\frac{1}{b_1}\tilde{\mathbf{V}}^{-1}\right]-\frac{1}{m}\mathds{1}_m\mathds{1}_m^\top\otimes\left[\left(\mathbf{M}(\xi)^{-1}+\tilde{\mathbf{R}}\right)^{-1}+\frac{v}{b_1}\tilde{\mathbf{V}}^{-1}\right]\right\}^{-1}\\
\quad &=& \left(\mathbb{I}_m-\frac{1}{m}\mathds{1}_m\mathds{1}_m^\top\right)\otimes\left[\left(\mathbf{M}(\xi)^{-1}+\tilde{\mathbf{R}}\right)^{-1}+\mathbf{V}_1^{-1}\right]^{-1}+\mathbf{C}_1,
\end{eqnarray*}
where $\mathbf{C}_1=\frac{1}{m}\mathds{1}_m\mathds{1}_m^\top\otimes\mathbf{V}_2$.
Then,
\begin{eqnarray*}
        \hspace{-1cm}\mathrm{MSE}_{\alpha}(\xi)&=&\mathbb{I}_f\otimes\left(\mathbb{I}_m-\frac{1}{m}\mathds{1}_m\mathds{1}_m^\top\right)\otimes\left[\left(\mathbf{M}(\xi)^{-1}+\tilde{\mathbf{R}}\right)^{-1}+\mathbf{V}_1^{-1}\right]^{-1} \\
        &+&\left(\mathbb{I}_f-\frac{1}{f}\mathds{1}_f\mathds{1}_f^\top\right)\otimes\frac{1}{m}\mathds{1}_m\mathds{1}_m^\top\otimes\left[\left(\mathbf{M}(\xi)^{-1}+\tilde{\mathbf{R}}\right)^{-1}+\mathbf{V}_2^{-1}\right]^{-1}+\mathbf{C}_2,
\end{eqnarray*}
where $\mathbf{C}_2=\frac{1}{f}\mathds{1}_f\mathds{1}_f^\top\otimes\mathbf{C}_1$.

For $i=1,2$ it holds that
\begin{equation*}
    \left\{[\mathbf{M}(\xi)^{-1}+\tilde{\mathbf{R}}]^{-1}+\mathbf{V}_i^{-1}\right\}^{-1}=\mathbf{V}_i-\mathbf{V}_i\mathbf{S}_i^{-1}\mathbf{V}_i+\mathbf{V}_i\mathbf{S}^{-1}\left[\mathbf{M}(\xi)+\mathbf{S}_i^{-1}\right]^{-1}\mathbf{S}_i^{-1}\mathbf{V}_i.
\end{equation*}
Thus,
\begin{eqnarray}
\mathrm{tr}\left[\mathrm{MSE}_{\alpha}(\xi)\left(\mathbb{I}_K\otimes\mathcal{L}\right)\right]&=&f(m-1)\mathrm{tr}\left\{\left[\mathbf{M}(\xi)+\mathbf{S}_1^{-1}\right]^{-1}\mathbf{S}_1^{-1}\mathbf{V}_1\mathcal{L}\mathbf{V}_1\mathbf{S}_1^{-1}\right\}\nonumber \\
&&+(f-1)\mathrm{tr}\left\{\left[\mathbf{M}(\xi)+\mathbf{S}_2^{-1}\right]^{-1}\mathbf{S}_2^{-1}\mathbf{V}_2\mathcal{L}\mathbf{V}_2\mathbf{S}_2^{-1}\right\}+const_2,\label{msetrbcs1}
\end{eqnarray}
where
\begin{equation*}
    const_2=f(m-1)\mathrm{tr}\left[\left(\mathbf{V}_1-\mathbf{V}_1\mathbf{S}_1^{-1}\mathbf{V}_1\right)\mathcal{L}\right]+(f-1)\mathrm{tr}\left[\left(\mathbf{V}_2-\mathbf{V}_2\mathbf{S}_2^{-1}\mathbf{V}_2\right)\mathcal{L}\right]+\mathrm{tr}\left(\mathbf{V}_2\mathcal{L}\right).
\end{equation*}

For matrix $\mathbf{T}=\mathbb{I}_K-\frac{1}{K}\mathds{1}_K\mathds{1}_K^\top$ it holds that 
\begin{equation*}
\mathbf{T}\left[\mathbb{I}_f\otimes\left(\mathbb{I}_m-\frac{1}{m}\mathds{1}_m\mathds{1}_m^\top\right)\right]=\mathbb{I}_f\otimes\left(\mathbb{I}_m-\frac{1}{m}\mathds{1}_m\mathds{1}_m^\top\right),
\end{equation*}
\begin{equation*}
\mathbf{T}\left[\left(\mathbb{I}_f-\frac{1}{f}\mathds{1}_f\mathds{1}_f^\top\right)\otimes\frac{1}{m}\mathds{1}_m\mathds{1}_m^\top\right]=\left(\mathbb{I}_f-\frac{1}{f}\mathds{1}_f\mathds{1}_f^\top\right)\otimes\frac{1}{m}\mathds{1}_m\mathds{1}_m^\top,
\end{equation*}
and $\left(\mathbf{T}\otimes \mathcal{L}\right)\mathbf{C}_2=\mathbf{0}$.
Thus,
\begin{eqnarray}
\mathrm{tr}\left[\mathrm{MSE}_{\theta}(\xi)\left(\mathbb{I}_K\otimes\mathcal{L}\right)\right]&=&f(m-1)\mathrm{tr}\left\{\left[\mathbf{M}(\xi)+\mathbf{S}_1^{-1}\right]^{-1}\mathbf{S}_1^{-1}\mathbf{V}_1\mathcal{L}\mathbf{V}_1\mathbf{S}_1^{-1}\right\}\nonumber \\
&&+(f-1)\mathrm{tr}\left\{\left[\mathbf{M}(\xi)+\mathbf{S}_2^{-1}\right]^{-1}\mathbf{S}_2^{-1}\mathbf{V}_2\mathcal{L}\mathbf{V}_2\mathbf{S}_2^{-1}\right\}+const_3,\label{msetrbcs2}
\end{eqnarray}
where
\begin{equation*}
    const_3=f(m-1)\mathrm{tr}\left[\left(\mathbf{V}_1-\mathbf{V}_1\mathbf{S}_1^{-1}\mathbf{V}_1\right)\mathcal{L}\right]+(f-1)\mathrm{tr}\left[\left(\mathbf{V}_2-\mathbf{V}_2\mathbf{S}_2^{-1}\mathbf{V}_2\right)\mathcal{L}\right].
\end{equation*}
The constant terms $const_2$ and $const_3$ have no influence on designs, and weighted \textit{A}-criteria \eqref{wacrappr} and \eqref{wacrapprdef} differ only by these constants from \eqref{msetrbcs1} and \eqref{msetrbcs2}, respectively.

\end{document}